\documentclass[10pt,aps,prd,twocolumn,superscriptaddress,nofootinbib,nobibnotes,longbibliography,floatfix]{revtex4-2}

\usepackage{calligra}
\usepackage{graphicx,mathtools,amssymb,amsmath,amsthm,amsfonts,epsfig,epsf,multirow}
\usepackage[outdir=./]{epstopdf}
\usepackage[dvipsnames]{xcolor}
\usepackage[linktocpage]{hyperref}
\hypersetup{
    colorlinks=true,
    linkcolor=blue,
    filecolor=magenta,      
    urlcolor=BrickRed,
    citecolor=BrickRed,
}
\hypersetup{pdfstartview=}	
\usepackage{tensor}	
\usepackage{csquotes}
\usepackage{mathrsfs}
\usepackage{float}

\usepackage{subfig}
\newcommand{\PRLsection}[1]{\noindent{\bf\emph{#1.---}}}

\usepackage{graphicx}
\usepackage{amsmath}
\usepackage{autobreak}
\usepackage{float}

\newcommand{\rom}[1]{\MakeUppercase{\romannumeral #1}}

\usepackage{multirow}
\usepackage{color, colortbl}

\usepackage[normalem]{ulem}

\begin{document}
\title{Spherical collapse in scalar-Gauss-Bonnet gravity:\\taming ill-posedness with a Ricci coupling}

\author{Farid~Thaalba}
\address{Nottingham Centre of Gravity,
Nottingham NG7 2RD, United Kingdom}
\address{School of Mathematical Sciences, University of Nottingham,
University Park, Nottingham NG7 2RD, United Kingdom}

\author{Miguel~Bezares}
\address{Nottingham Centre of Gravity,
Nottingham NG7 2RD, United Kingdom}
\address{School of Mathematical Sciences, University of Nottingham,
University Park, Nottingham NG7 2RD, United Kingdom}

\author{Nicola~Franchini}
\affiliation{Université Paris Cit\'e, CNRS, Astroparticule et Cosmologie,  F-75013 Paris, France}
\affiliation{CNRS-UCB International Research Laboratory, Centre Pierre Binétruy, IRL2007, CPB-IN2P3, Berkeley, CA 94720, USA}

\author{Thomas~P.~Sotiriou}
\address{Nottingham Centre of Gravity,
Nottingham NG7 2RD, United Kingdom}
\address{School of Mathematical Sciences, University of Nottingham,
University Park, Nottingham NG7 2RD, United Kingdom}
\address{School of Physics and Astronomy, University of Nottingham,
University Park, Nottingham NG7 2RD, United Kingdom}

\begin{abstract}
We study  spherical collapse of a scalar cloud in scalar-Gauss-Bonnet gravity --- a theory in which black holes can develop scalar hair if they are in a certain mass range. We show that an additional quadratic coupling of the scalar field to the Ricci scalar can mitigate loss of hyperbolicity problems that have plagued previous numerical collapse studies and instead lead to well-posed evolution. This suggests that including specific additional interactions  can be a successful strategy for tackling well-posedness problems in effective field theories of gravity  with nonminimally coupled scalars. Our simulations also show that spherical collapse leads to black holes with scalar hair when their mass is below a mass threshold and above a minimum mass bound and that above the mass threshold the collapse leads to black holes without hair, in line with results in the static case and perturbative analyses. For masses below the minimum mass bound we find that the scalar cloud smoothly dissipates, leaving behind flat space.  
\end{abstract}

\maketitle

%----------------------------------------------------------------------------------------------------
\PRLsection{Introduction}
%-------------------------------------------------------------------------------------------------
Astrophysical phenomena in which gravity is at its strongest regime, such as the collapse of a compact object or the merger of coalescing binaries, cannot be adequately described with perturbative techniques. To model them in General Relativity (GR), one instead needs to formulate Einstein's equations as an initial value problem (IVP) and solve them numerically. Testing gravity in its nonlinear regime, to search for new physics beyond GR or the Standard Model of particle physics that could manifest, requires formulating IVPs and performing numerical simulations in such beyond-GR scenarios. To obtain predictions a well-posed IVP formulation should exist, {\it i.e.}, the solution would be unique and depend continuously on the initial conditions. 

A major obstacle is that in many interesting such scenarios, it is not obvious how to obtain a well-posed formulation of the IVP because the new physics drastically changes the structure of the field equations as partial differential equations. Scalar fields provide a characteristic example. No-hair theorems \cite{Hawking:1972qk,Bekenstein:1995un,Sotiriou:2011dz,Hui:2012qt,Sotiriou:2015pka,Herdeiro:2015waa} dictate that scalars cannot leave an imprint on black holes (BHs) in most cases. All the known counterexamples ({\em e.g.}~\cite{Mignemi:1992nt,Kanti:1995vq,Sotiriou:2013qea,Sotiriou:2014pfa,Doneva:2017bvd,Silva:2017uqg,Antoniou:2017hxj}) require coupling the scalar to the Gauss-Bonnet (GB) invariant,  $\mathcal{G}=R^{\mu \nu \rho \sigma} R_{\mu \nu \rho \sigma}-4 R^{\mu \nu} R_{\mu \nu}+R^{2}$, where $R_{\mu \nu \rho \sigma}$, $R_{\mu \nu}$, and $R$  being the Riemann tensor, Ricci tensor, and Ricci scalar respectively.  Einstein's equations are quasi-linear, {\em i.e.}~linear in the second derivatives of the metric, and this is a key property in what regards their well-posedness as an IVP. In contrast, $\mathcal{G}$ is clearly quadratic in the curvature and hence quadratic in the second derivatives of the metric. This places well-posedness in jeopardy.

One can circumvent this problem by working perturbatively in the coupling constants that control the deviations from GR. Assuming that the solutions are continuously connected to GR as the coupling goes to zero, one can generate solutions order by order~\cite{Benkel:2016rlz,Benkel:2016kcq,Okounkova:2017yby,Witek:2018dmd,Okounkova:2018abo,Okounkova:2019dfo}. One disadvantage of this approach is that secular growth can drive it out of its range of validity~\cite{GalvezGhersi:2021sxs}. Another is that the perturbative treatment in the coupling is not suitable for capturing effects that involve nonlinearities in the new fields.

This last point is clearly demonstrated in the phenomenon of scalarization \cite{Doneva:2022ewd}, first introduced in Ref.~\cite{Damour:1993hw} for neutron stars. It was shown more recently that scalarization can affect BHs as well if a scalar field exhibits a suitable coupling with the GB invariant $\mathcal{G}$ \cite{Silva:2017uqg,Doneva:2017bvd}. Scalarization can be understood as a linear tachyonic instability of the scalar that is eventually nonlinearly quenched. The instability occurs in GR spacetimes describing compact objects, it is controlled by interactions that are quadratic in the scalar \cite{Andreou:2019ikc} and, for BHs, it appears below a mass \cite{Silva:2017uqg,Doneva:2017bvd} or above a spin \cite{Dima:2020yac,Herdeiro:2020wei} threshold. The endpoint is a compact object with a nontrivial scalar configuration, whose properties are determined by nonlinear interactions of the scalar \cite{Silva:2018qhn,Macedo:2019sem,Antoniou:2021zoy}. Hence, the dynamics of theories exhibiting scalarization cannot be fully captured working perturbatively in the coupling constant \cite{Palenzuela:2013hsa,Silva:2020omi,East:2021bqk,Doneva:2022byd,Elley:2022ept}. 

Indeed, the study of the IVP for scalars nonminimally coupled to gravity, and in particular for the broad class of theories that lead to second-order equations described by the Horndeski action \cite{Horndeski:1974wa}, has received a lot of attention recently \cite{Papallo:2017qvl,Papallo:2017ddx,Ripley:2019hxt,Ripley:2019irj,Ripley:2020vpk,Bezares:2020wkn,Figueras:2020dzx,Figueras:2021abd,Bernard:2019fjb}. It was established in \cite{Kovacs:2020ywu} that an appropriate formulation exists that renders the IVP well-posed in these theories in the weakly coupled regime, {\it i.e.}, when the nonminimal couplings of the scalar remain small. Numerical studies, restricted so far to theories where the scalar couples only to the GB invariant (see \cite{Liu:2022fxy} for an exception), have verified this result for small values of the coupling constant that controls this coupling \cite{East:2020hgw,East:2021bqk,Corman:2022xqg,AresteSalo:2022hua}. For larger values of the coupling constant however well-posedness is eventually lost, as the equations tend to change character from hyperbolic to elliptic in a region of spacetime \cite{Ripley:2019hxt,Ripley:2019irj,Ripley:2020vpk,East:2020hgw,East:2021bqk,Corman:2022xqg,AresteSalo:2022hua,R:2022hlf}. 

Viewing these theories as nonlinear effective field theories (EFTs), and hence as products of some truncation of a more ``fundamental" theory, a promising strategy could be to try to employ a method inspired by viscous relativistic hydrodynamics to render them predictive~\cite{Cayuso:2017iqc}. How to import such a method to gravity theories is currently being explored~\cite{Allwright:2018rut,Cayuso:2020lca,Franchini:2022ukz,Cayuso:2023aht,Lara:2021piy,Bezares:2021dma,Bezares:2021yek,Cayuso:2023aht,Gerhardinger:2022bcw}. An interesting alternative would be to study whether additional couplings of the scalar, which one could expect to be there in an EFT, could actually improve the theories behaviour and lead to well-posed evolution. Indeed, it has been shown in Ref.~\cite{Antoniou:2022agj} that an additional coupling of the type $\phi^2 R$ leads to an improvement of the hyperbolic nature of the equations for linear perturbations in scalarization theories that contain $\phi^2 \mathcal{G}$ couplings. Interestingly, that same curvature interaction, $\phi^2 R$, has been shown to resolve radial stability problems for scalarized BHs \cite{Antoniou:2021zoy,Antoniou:2022agj}, to help evade binary pulsar constraints by suppressing neutron star scalarization~\cite{Ventagli:2021ubn}, and to render GR a cosmological attractor without the need for the fine-tuning of the initial conditions~\cite{Antoniou:2020nax}, thereby making scalarization models compatible with late time cosmological dynamics.

Motivated by the above,  we consider here a theory with quadratic scalar couplings to both the GB invariant and the Ricci scalar and study gravitational collapse of a scalar cloud. We show that the Ricci coupling does indeed improve significantly the dynamical behaviour of the theory and allows it to be predictive for values of the GB coupling that otherwise would have yielded an ill-posed IVP. Our numerical simulations allow us to track the formation of a scalarized BH for suitable initial data. For data that would have led to a formation of a BHs with mass smaller than the existence threshold of the theory we instead see the scalar cloud smoothly dispersing.   

%----------------------------------------------------------------------------------------------------
\PRLsection{The theory}
%----------------------------------------------------------------------------------------------------
We consider the following action: 
\begin{align}
    \label{eq:action}
    S= \frac{1}{16 \pi} \int \mathrm{d}^{4} x \sqrt{-g}\left[R+X-\left(\frac{\beta}{2}R - \alpha \mathcal{G}\right)\frac{\phi^2}{2}\right],
\end{align}
where $g = \text{det}(g_{\mu \nu})$, $X=-\nabla_\mu\phi\nabla^\mu\phi/2$,
and we use units of $G=c=1$. The parameter
$\beta$ is a dimensionless coupling constant while $\alpha$ is of dimension length squared.

Varying action \eqref{eq:action} with respect to the metric yields,
\begin{align}
    \label{eq:einstein}
    G_{\mu \nu} &= T^{\phi}_{\mu \nu}~,
\end{align}
where
\begin{align}
    T_{\mu \nu}^{\phi}&= -\frac{1}{4} g_{\mu \nu}(\nabla \phi)^{2}+\frac{1}{2} \nabla_{\mu} \phi \nabla_{\nu} \phi \nonumber \\  
    & -\frac{\alpha}{2 g} g_{\mu(\rho} g_{\sigma) \nu} \epsilon^{\kappa \rho \alpha \beta} \epsilon^{\sigma \gamma \lambda \tau} R_{\lambda \tau \alpha \beta} \nabla_{\gamma} \nabla_{\kappa} \phi^2 \nonumber \\ 
    & +\frac{\beta}{4}\left(G_{\mu \nu} + g_{\mu \nu} \Box - \nabla_{\mu}\nabla_{\nu}\right)\phi^2~.
    \label{eq:scalar_stressenergy}
\end{align}
and with respect to the scalar fields, gives
\begin{align}
    \label{eq:scalar}
    \Box \phi &= \frac{\beta}{2}R - \alpha \mathcal{G}~,
\end{align}
where $\Box \coloneqq \nabla_{\mu}\nabla^{\mu}$, $G_{\mu\nu}$ is the Einstein tensor, and $\epsilon^{\sigma \gamma \lambda \tau}$ is the Levi-civita totally anti-symmetric tensor density.

%----------------------------------------------------------------------------------------------------
\PRLsection{Evolution in spherical symmetry}
%----------------------------------------------------------------------------------------------------
In this paper, we focus on the study of the IVP in spherical symmetry. We follow closely the prescription given in~\cite{Ripley:2019irj}.
To this end, we consider a spherically-symmetric background, with the following ansatz in polar coordinates $(t,r,\theta,\varphi)$
\begin{align}
    \label{eq:metric_ansatz}
    \mathrm{d}s^2 = -\mathrm{e}^{2 A(t, r)} \mathrm{d} t^{2}+\mathrm{e}^{2 B(t, r)} \mathrm{d} r^{2}+r^{2} \mathrm{~d} \Omega^{2}~,
\end{align}
and same symmetries for the scalar field $\phi = \phi(t,r)$. 
We introduce new variables
\begin{align}
P(t, r) \coloneqq \mathrm{e}^{-A+B} \partial_{t} \phi, \qquad Q(t, r) \coloneqq \partial_{r} \phi.
\end{align}
With this choice of variables and the ansatz of Eq.~\eqref{eq:metric_ansatz}, the system~\eqref{eq:einstein}--\eqref{eq:scalar} is reduced to three time-evolution equations for $\phi$, $P$ and $Q$, and two radial constraints for $A$ and $B$.

This system of equations might not always admit a well-posed IVP. We will say that the system is well-posed if there exists a unique solution that depends continuously on its initial data. {This will be the case if the system is strongly hyperbolic, {\em i.e.}~its principal part (the highest derivative terms) is diagonalizable with real eigenvalues \cite{doi:10.1137/1.9780898719130, Sarbach:2012pr}.}

In numerical considerations, the characteristic speeds constitute a useful diagnostic tool as they convey information regarding the character of the evolution equations. In spherical symmetry, they are given by \cite{Ripley:2019irj}
\begin{align}
    \label{eq:char_speeds}
    c_{\pm} = \frac{1}{2} \left(\text{Tr}(\mathcal{C}) \pm \sqrt{\mathcal{D}}\right),
\end{align}
where $\mathcal{C}$, and $\mathcal{D}$ depend on the functions $A$, $B$, $\phi$, $P$, and $Q$, and their derivation can be found in the Appendix.
The system is hyperbolic if $\mathcal{D}>0$, otherwise, the system is elliptic ($\mathcal{D}<0$) or parabolic ($\mathcal{D}=0$).

%----------------------------------------------------------------------------------------------------
\PRLsection{Initial data}
%----------------------------------------------------------------------------------------------------
As for initial data (ID), we are free to specify the values of $\phi$ and $P$ at $t=0$, whereas by definition we must have $Q(0,r) = \partial_r\phi(0,r)$.  
We use two types of initial data, type-\rom{1} ID is a static Gaussian pulse
\begin{align}
\label{eq:initialdata1}
    \phi(0,r) &= a_0 \exp\left[-\left(\frac{r-r_0}{w_0}\right)^2\right], \quad P(0,r) = 0,
\end{align}
while type-\rom{2} ID is an approximately in-going pulse
\begin{align}
\label{eq:initialdata2}
    \phi(0,r) &= a_0 \left(\frac{r}{w_{0}}\right)^2\exp\left[-\left(\frac{r-r_0}{w_0}\right)^2\right],\\
    P(0,r) &= -\frac{1}{r}\phi(0,r) - Q(0,r),
\end{align}
where $a_{0}$, $r_{0}$ and $w_0$ are constants.

%----------------------------------------------------------------------------------------------------
\PRLsection{Numerical implementation}
%----------------------------------------------------------------------------------------------------
We employ a fully constrained evolution scheme. The equations are discretized over the domain $[0,T] \times [0, R]$ for some choice of $R$, and $T$. For a given resolution $N$, the radial step $\Delta r$ is given by $\Delta r = R/N$, and the time step is defined by setting the Courant parameter to $\lambda = 0.25$ i.e., $\Delta t = \lambda \Delta r$. To impose regularity at the centre we take a staggered grid and perform the following expansion 
\begin{align}\label{eq:Taylor_phi}
    \phi(t,r) &= \phi_{0}(t) + \phi_{2}(t)r^2 + \phi_{4}(t)r^4 + \mathcal{O}(r^6)~, \\
    A(t,r) &= A_{0}(t) + A_{2}(t)r^2 + A_{4}(t)r^4 + \mathcal{O}(r^6)~, \\
    B(t,r) &= B_{0}(t) + B_{2}(t)r^2 + B_{4}(t)r^4 + \mathcal{O}(r^6)~, \label{eq:Taylor_B}
\end{align}
and solve for $A_0, A_2, A_4, B_2, B_4$ by using the initial data. 
Moreover, the boundary conditions at $r=0$ require that $\partial_r A = 0$ and $\partial_r B = 0$. One can see from the equations that by choosing $B_0 = 0$ this is automatically satisfied. Without such treatment at the origin, it was only possible to evolve the system for a small subset of coupling constants. At the outer boundary of the domain, we impose approximately outgoing boundary conditions for $\phi, Q, P$, while the metric functions are completely specified by solving the constraint. 
Given the initial data, we solve the constraint equations for the metric functions on the discretized domain $[0,R]$ using a fourth-order Runge-Kutta (RK4) scheme. This scheme requires knowledge of the value of $\phi, Q$, and $P$ at intermediate virtual points and we obtain such information by employing a fifth-order Lagrangian interpolator. Once we have the solution, we set $A(t,r) \rightarrow A(t,r)-A(t,R)$ by utilizing the remaining gauge freedom. This guarantees that at the outer boundary of the domain, the time function $t$ measures the proper time of a static observer. After obtaining the metric functions, we integrate the variables $\phi, Q$, and $P$ in time using the method of lines with an RK4 scheme and a sixth-order Kreiss-Olliger (KO) dissipation term. We use a fourth-order finite differences operator satisfying summation by parts to discretize radial derivatives \cite{Calabrese:2003vx}. At the initial step we compute the initial Misner-Sharp mass~\cite{Misner:1964je} as $M = R\left(1 - \mathrm{e}^{-2B(0,R)}\right)/2$, which we use for the rescaling of the output. We keep track of the formation of an apparent horizon by checking the largest radius for which $\exp(A-B)$ falls below a chosen tolerance, indicating the formation of a BH.

\begin{figure}[t]
     \centering
     \includegraphics[width=1.\linewidth]{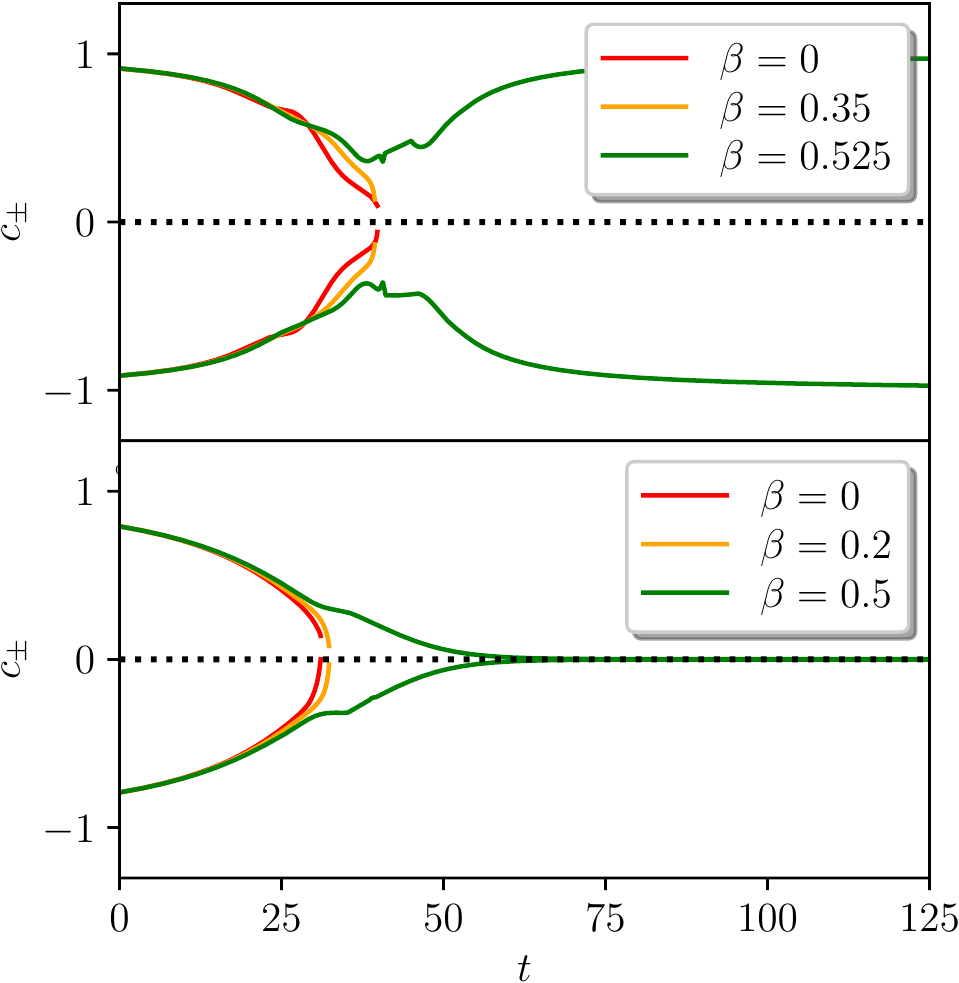}
     \caption{ Both plots are for $\alpha/M^2 = 0.25$ and type-\rom{2} ID with $r_0 = 25$, $w_0 = 6$. The plot on the top is for $a_{0}=0.01$, and the one on the bottom is for $a_{0}=0.016$. We show the maximum of $c_{-}$ (solid lines) and the minimum of $c_{+}$ (dashed lines) in space for different values of $\beta$. In the top panel, for $\beta=0, 0.35$ the characteristic speeds approach zero and will eventually cross it and change sign. Therefore, the equations change character and we cannot follow the evolution further, but for $\beta=0.525$ the characteristic speeds do not cross zero and the end state of the evolution is flat geometry. In the bottom panel, we observe the formation of an apparent horizon for $\beta=0.5$. The curves indeed asymptote to zero, unlike the $\beta=0, 0.2$ cases.}
    \label{fig:speeds}
\end{figure}

%----------------------------------------------------------------------------------------------------
\PRLsection{Results}
%----------------------------------------------------------------------------------------------------
We first study how the inclusion of the Ricci term improves the hyperbolic nature of the evolution system for a specific value of $\alpha/M^2=0.25$. In Fig.~\ref{fig:speeds} we plot how the maximum of $c_{-}$ and the minimum of $c_{+}$ vary with time for different values of $\beta$, type-\rom{2} ID with $r_0 = 25$, $w_0 = 6$ and for two different amplitudes $a_0 = {0.01, 0.016}$. A vanishing or rather small value of $\beta$ leads to the formation of a naked elliptic region (NER), {\em i.e.}~a region of spacetime where the characteristic speeds change sign signalling that the character of the equations has changed from hyperbolic to elliptic without being shielded by an apparent horizon~\cite{Ripley:2019aqj}. On the other hand, for a sufficiently large coupling to the Ricci scalar, the plot suggests that this coupling heals the problem and allows for the evolution to continue for later times. 

\begin{figure}[t]
    \centering
    \includegraphics[width=1.\linewidth]{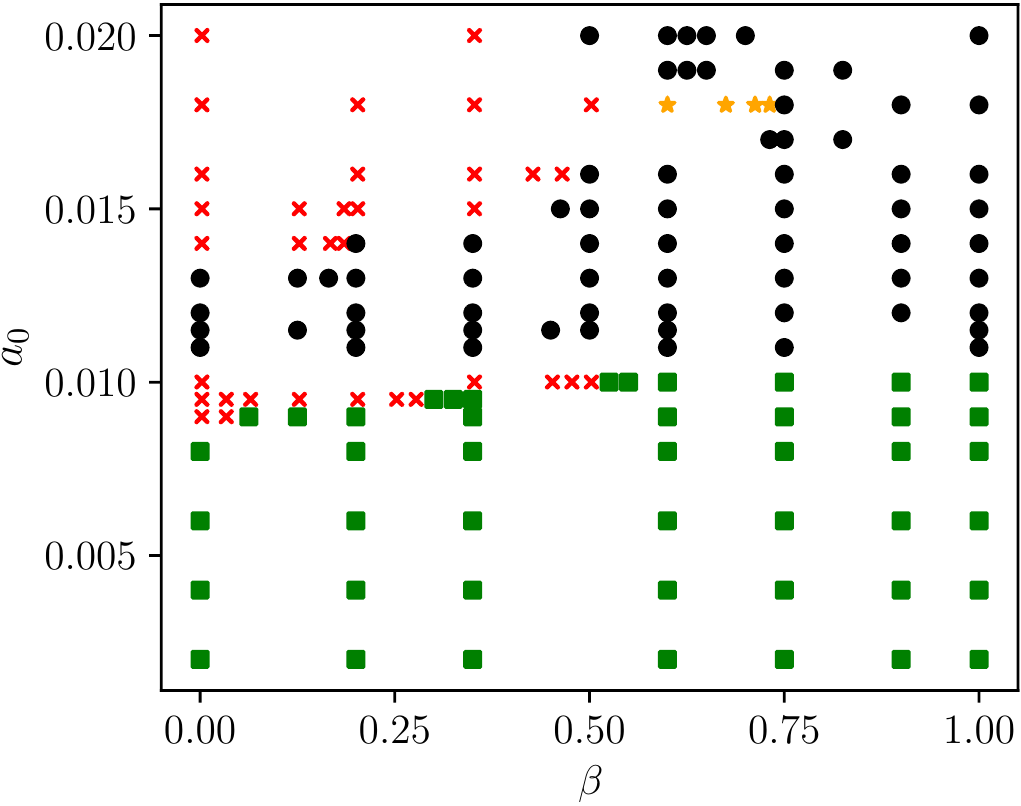}
    \caption{Scatter plot for $\alpha/M^2 = 0.25$ and type-\rom{2} ID with $r_0 = 25$, $w_0 = 6$, and a varying amplitude $a_0$. The end state of the evolution is indicated by green squares for flat spacetimes, red crosses for NERs, and black dots for BHs. Orange stars denote cases for which it is difficult to conclude if the end state is a BH or a NER (in polar coordinates). For large enough $\beta$ the Ricci coupling ``cures'' the loss of hyperbolicity for both flat space and BH end states.}
    \label{fig:main}
\end{figure}

Motivated by these results, we explore the parameter space more thoroughly.
We start with fixed $\alpha/M^2 = 0.25$ and type-\rom{2} ID with $r_0 = 25$, $w_0 = 6$, and varied the amplitude $a_0$ in the range $[0.2,2]\times10^{-2}$ and the Ricci coupling $\beta$ in the range $[0,1]$. For each of the cases considered, we establish whether the outcome of the evolution is flat space, NER, or the development of an apparent horizon. Outcomes are summarized in Fig.~\ref{fig:main}. For $\beta=0$ and low enough initial amplitudes, the theory is predictive, and the outcome is flat spacetime. For certain larger amplitudes an apparent horizon forms. However, we encounter a NER both in the transition between these two outcomes and when we increase the amplitude further. Remarkably, increasing the value of $\beta$ eventually removes the NER in all cases.

On a few occasions, labelled by an orange star in Fig.~\ref{fig:main}, it was not possible to conclude whether the outcome is a formation of an apparent horizon or a NER due to the steep gradients in the metric sector. This is due to the fact that polar coordinates are not horizon penetrating and it does not affect our previous conclusions regarding the effect of the Ricci coupling. We have performed additional simulations with different values of $\alpha/M^2$ and for the two different types of ID that confirm the positive effect this coupling has on hyperbolicity in the case of spherical collapse. A summary of the other simulations performed can be found in table \ref{tab:runs} in the Appendix~\ref{parspace}.

\begin{figure}[t]
    \centering
    \includegraphics[width=0.45\textwidth]{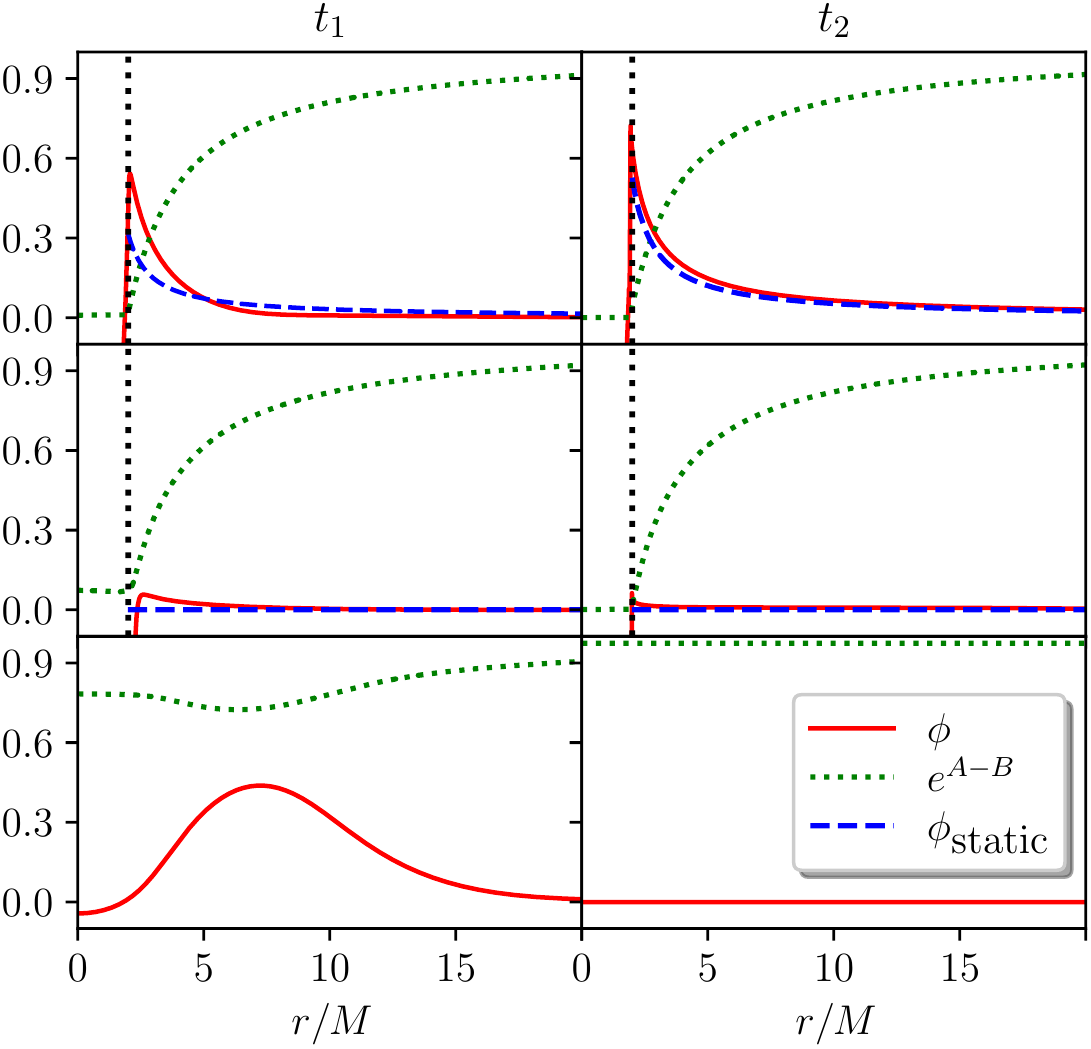}
    \caption{Scalar field and  $e^{A-B}$, which vanishes at the apparent horizon, as a function of radius for two different time instances for type-\rom{2} ID with fixed $r_{0}=25$, and $w_{0} = 6$. $\phi_{\text{static}}$ is the static solution. The black dotted line in the first two panels is the location of the apparent horizon. \textbf{Top panel}: is for $\alpha/M^2 = 0.75, \beta=1.5$ with $a_{0} = 0.015$. Snapshots with $\{t_1,t_2\}/M=\{22, 88\}$ show the formation of an apparent horizon where the quantity $e^{A-B} \rightarrow 0$, with the late time behavior maintaining a scalar profile with a $1/r$ fall off. The blue dashed curve shows the static solution for the corresponding $\alpha/M^2$ at each instance (as the mass might decrease during the evolution). \textbf{Middle panel}: is for $\alpha/M^2 = 0.25, \beta=1$ with $a_{0} = 0.02$ and $\{t_1, t_2\}/M=\{11, 28\}$. In this case, an apparent horizon forms but the late-time behavior does not support a scalar profile. \textbf{Bottom panel}: is for $\alpha/M^2 = 1.25, \beta=2$ with $a_{0} = 0.01$ and $\{t_1, t_2\}/M=\{39, 97\}$. Here, the end state is flat geometry for which the scalar field dissipates to infinity.}
    \label{fig:scalar_field}
\end{figure}

Since for large enough $\beta$ we can track the evolution, we delve a bit deeper into the properties of the end state. We present three representative cases in Fig.~\ref{fig:scalar_field}, where we show an early and a late snapshot of the evolution of the scalar field $\phi$ and the metric combination $\exp{(A-B)}$, which vanishes on the apparent horizon, for type-\rom{2} ID with fixed $r_{0}=25$, and $w_{0} = 6$ for three different cases. (The fact that $\exp{(A-B)}$ vanishes also at smaller radii, once an apparent horizon forms, is due to the use of coordinates that are not horizon-penetrating.) For each of them, the amplitude $a_0$ is chosen such that it would produce an apparent horizon in GR.
From the analysis of static solutions in Ref.~\cite{Antoniou:2021zoy}, we know that, for large enough values of $\beta$, BHs below some threshold mass and down to some minimum mass will exhibit hair. BHs over the threshold mass will have no hair. Below that minimum mass, the Schwarzschild metric is unstable and no hairy BHs appear to exist either. The parameters of the three panels of Fig.~\ref{fig:scalar_field}, top to bottom, have been chosen such that the mass associated with the ID corresponds to each of the three cases respectively. As can be seen from the plots, the end state of evolution is in agreement with the analysis of the static solutions. When applicable, we provide a comparison with a static and spherically symmetric profile, obtained by integrating numerically the equations~\eqref{eq:einstein}--\eqref{eq:scalar}, as described in~\cite{Antoniou:2021zoy}. Note that the lack of perfect overlap between the scalar profile and the static configuration is in part due to the limitations of using polar coordinates. Remarkably, in the case where the static limit predicts that no (stable) BH can exist (bottom panel), we find no impediment in the time evolution and the scalar field dissipates to infinity leaving a flat background.

%----------------------------------------------------------------------------------------------------
\PRLsection{Conclusions}
%----------------------------------------------------------------------------------------------------
We have investigated the effect of non-minimally coupling the scalar field quadratically to the Ricci scalar on the well-posedness of the IVP in sGB gravity, in the case of spherical collapse. Our results show that this additional coupling, for large enough values of the corresponding coupling constant, can mitigate against the formation of a NER, which signals a loss of hyperbolicity and plagued earlier numerical simulations. This demonstrates that including specific additional interactions --- other than those that are essential for having interesting phenomenology, such as BH hair --- can be a successful strategy for tackling well-posedness problems in EFTs of gravity  with nonminimally coupled scalars.

There are three important limitations to our results, which also highlight interesting future directions. The first one is spherical symmetry. It is likely that the coupling to the Ricci scalar might not be sufficient to cure ill-posedness in a less symmetric setup and that a broader range of couplings would need to be explored. $3+1$ simulations would also allow us to explore numerically the non-radial stability of scalarized black holes \cite{Kleihaus:2023zzs,Minamitsuji:2023uyb}.The second limitation is that we only considered the collapse of a scalar cloud. Work is already underway to generalize our results to the case of stellar collapse. The third limitation is that our numerical implementation involved coordinates that are not horizon penetrating, and hence are not ideal for probing the properties of the end state when the latter is a BH.

Interestingly, our simulations did allow us to confirm the expectations arising from combining static results and linear perturbations ({\em e.g.}~\cite{Silva:2017uqg,Doneva:2017bvd,Antoniou:2021zoy,Antoniou:2022agj,Doneva:2022ewd}: that spherical collapse will lead to BHs with scalar hair when their mass is below a mass threshold and above a minimum mass bound and that above the mass threshold collapse leads to BHs without hair. Remarkably, in simulations that would form a BH below the minimum mass bound, where stable BHs are not know to exist, the scalar cloud smoothly dissipated, leaving behind flat space. We are currently exploring all of these cases in more detail in a numerical implementation that uses horizon-penetrating coordinates, which should allow us to trace the evolution past the formation of an apparent horizon and probe the end state in more detail.

\acknowledgements
The authors would like to thank Luis Lehner for fruitful discussions about the numerical implementation of the problem. T.P.S. acknowledges partial support from
the STFC Consolidated Grants no. ST/T000732/1 and
no. ST/V005596/1.

\section*{appendix}
%----------------------------------------------------------------------------------------------------
\PRLsection{Characteristic speeds}\label{app:charspeed}
%----------------------------------------------------------------------------------------------------
Given a system of first-order partial differential equations $V_{I}(x, u, \partial u)=0$,\footnote{The introduction of the new variables $P$ and $Q$ allows for writing the EOM as a first-order system of PDEs, therefore, we focus on such systems.} with the index $I$ counting the number of equations, $x^{\mu}$ being the spacetime coordinates, and $u^{J}$ denoting the field content. The principal symbol is defined as follows \cite{Ripley:2022cdh,Ripley:2019irj,Kovacs:2020ywu,Sarbach:2012pr}
\begin{align}
    \mathcal{P}_{IJ}(\xi) \coloneqq \mathcal{P}_{IJ}^{\mu} \xi_{\mu}= \frac{\partial V_{I}}{\partial(\partial_{\mu}u^J)}\xi_{\mu},
\end{align}
for a given covector $\xi_{\mu}$. The covector $\xi_{\mu}$ is called characteristic if it satisfies the characteristic equation 
\begin{align}
    \label{eq:char_eqn}
    \text{det}(\mathcal{P}_{IJ}(\xi)) = 0.
\end{align}
The system of equations is well-posed if all solutions of the characteristic equation are real. In spherical symmetry, the principal symbol matrix can be written as \cite{Ripley:2019irj}
\begin{align}
    \begin{aligned}
        \mathcal{P}(\xi)=\left(\begin{array}{cc}
        \mathcal{A} \xi_t+\mathcal{B} \xi_r & \mathcal{Q} \xi_r \\
        \mathcal{R} \xi_r & \mathcal{S} \xi_r
        \end{array}\right), 
    \end{aligned}
\end{align}
where
\begin{align}
    \begin{aligned}
\mathcal{A} & \coloneqq\left(\begin{array}{ll}
\partial E_{Q} / \partial\left(\partial_t Q\right) & \partial E_{Q} / \partial\left(\partial_t P\right) \\
\partial E_{P} / \partial\left(\partial_t Q\right) & \partial E_{P} / \partial\left(\partial_t P\right)
\end{array}\right), \\
\mathcal{B} & \coloneqq\left(\begin{array}{ll}
\partial E_{Q} / \partial\left(\partial_r Q\right) & \partial E_{Q} / \partial\left(\partial_r P\right) \\
\partial E_{P} / \partial\left(\partial_r Q\right) & \partial E_{P} / \partial\left(\partial_r P\right)
\end{array}\right), \\
\mathcal{Q} & \coloneqq\left(\begin{array}{ll}
\partial E_{Q} / \partial\left(\partial_r A\right) & \partial E_{Q} / \partial\left(\partial_r B\right) \\
\partial E_{P} / \partial\left(\partial_r A\right) & \partial E_{P} / \partial\left(\partial_r B\right)
\end{array}\right), \\
\mathcal{R} & \coloneqq\left(\begin{array}{ll}
\partial C_{A} / \partial\left(\partial_r Q\right) & \partial C_{A} / \partial\left(\partial_r P\right) \\
\partial C_{B} / \partial\left(\partial_r Q\right) & \partial C_{B} / \partial\left(\partial_r P\right)
\end{array}\right), \\
\mathcal{S} & \coloneqq\left(\begin{array}{ll}
\partial C_{A} / \partial\left(\partial_r A\right) & \partial C_{A} / \partial\left(\partial_r B\right) \\
\partial C_{B} / \partial\left(\partial_r A\right) & \partial C_{B} / \partial\left(\partial_r B\right)
\end{array}\right) .
\end{aligned}
\end{align}
where $E_Q=0$, $E_P=0$ are the evolution equations, and $C_A=0$, $C_B=0$ are the constraint equations. The characteristic speed in spherical symmetry is defined as $c \coloneqq -\frac{\xi_{t}}{\xi_{r}}$, 
if $\xi_{\mu}$ satisfies the characteristic equation then the corresponding values of the characteristic speeds are given by
\begin{align}
    \label{eq:char_speeds}
    c_{\pm} = \frac{1}{2} \left(\text{Tr}(\mathcal{C}) \pm \sqrt{\mathcal{D}}\right),
\end{align}
where, 
\begin{align}
    \mathcal{D} & \coloneqq \text{Tr}(\mathcal{C})^2 - 4\text{Det}(\mathcal{C}), \\ 
    \mathcal{C} & \coloneqq \mathcal{A}^{-1} \cdot\left(\mathcal{B}-\mathcal{Q} \cdot \mathcal{S}^{-1} \cdot \mathcal{R}\right) .
\end{align}

\begin{figure}
    \centering
    \includegraphics[width=0.45\textwidth]{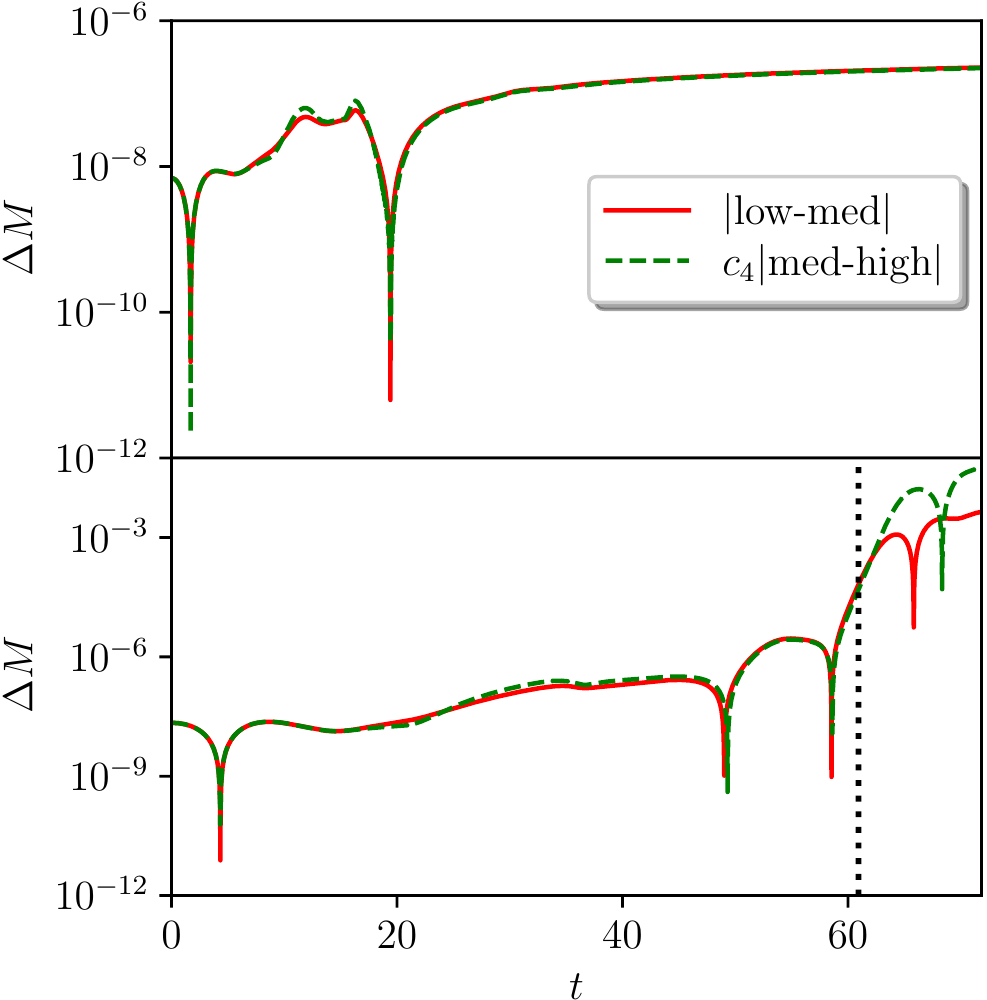}
    \caption{In both panels, we show the convergence of the Misner-Sharp mass where we plot the absolute difference between the low and medium resolutions, and between the medium and high resolutions for type \rom{1}-ID with $r_0 = 25$, $w_0 = 6$. The latter is rescaled by a factor of $c_{4}=16$ for both plots showing fourth-order convergence.  In the top panel, we have $a_0 = 0.3$, and $\alpha/M^2 = 1.25$, $\beta = 0.5$ with the end state being flat spacetime. In the bottom panel, we have $a_0 = 0.4$, and $\alpha/M^2 = 0.025$, $\beta = 0.9$, which forms an apparent horizon indicated by the black dotted line.}
    \label{fig:conv_test}
\end{figure}

%----------------------------------------------------------------------------------------------------
\PRLsection{Code validation}\label{codevalid}
%----------------------------------------------------------------------------------------------------
Here, we present the convergence tests to validate the simulations we have performed. We present two cases, one for which the end state is a flat background, and another where we observe the formation of an apparent horizon. We use three different resolutions $\{\Delta r_{\text{Low}}, \Delta r_{\text{Mid}}, \Delta r_{\text{High}}\} = \{0.0625, 0.03125, 0.015625\}$. In Fig.~\ref{fig:conv_test} we display the convergence behavior of the Misner-Sharp mass. In both panels, we show fourth-order convergence up until the formation of the apparent horizon (in the second panel) which is expected.  

%----------------------------------------------------------------------------------------------------
\PRLsection{Exploring the parameter space}\label{parspace}
%----------------------------------------------------------------------------------------------------
In table~\ref{tab:runs}, we summarize some additional simulations we performed. For each case, we specify what type of ID we used, the amplitude of the pulse $a_0$ (we fix for all the cases $r_0 = 25$ and $w_0 = 6$), the value of the coupling constants $\alpha/M^2$ and $\beta$, the Misner-Sharp mass $M$ at $t=0$ that we use to rescale lengths, and finally the outcome of the simulation. Among the outcomes, we distinguish the following cases. We dub \textit{flat} all the cases in which the scalar field disperses to infinity; \textit{NER} when the discriminant $\mathcal{D}$ becomes negative, signalling the appearance of a NER; \textit{BH} and \textit{sBH} when we can trace the formation of an apparent horizon from the condition $\exp(A-B) \to 0$. In the former, the scalar field approaches zero in the late time stages of the evolution, while in the latter, it is consistent with stationary scalarized BHs. 

Finally, we note that in some cases, which encompass large values of $\alpha/M^2$ and $\beta$, we encountered difficulty in imposing the boundary conditions at the centre. A solution to this problem was to increase the order of the Taylor expansion at the centre to select accurate boundary conditions for $A$ and $B$ as shown in equations~\eqref{eq:Taylor_phi}--\eqref{eq:Taylor_B}. We also noticed that increasing the strength of the KO dissipation and decreasing the order of the stencils of finite differences from fourth to second order mitigated this effect. 

\begin{table}[h!]
    \centering
    \begin{tabular}{cc|cc|c|c}
      \multicolumn{2}{c|}{ID} & \multicolumn{2}{c|}{Coupling constants} & \multirow{2}*{$M$} & \multirow{2}*{Outcome}\\
       type & $a_0$ & $\alpha/M^2$ & $\beta$ & \\
       \hline
        \rom{1} & 0.1 & 0.0 & 0.0 & 0.1693 & flat\\
        \rom{1} & 0.1 & \{0.75, 1.25\} & 0.0 & 0.1693 & flat\\
        \rom{1} & 0.1 & \{0.75, 1.25\} & 0.5 & 0.1692 & flat\\
        \rom{1} & 0.1 & \{0.75, 1.25\} & 1.0 & 0.1694 & flat\\
        \rom{1} & 0.1 & \{0.75, 1.25\} & 1.5 & 0.1699 & flat\\
        \rom{1} & 0.1 & \{0.75, 1.25\} & 2.0 & 0.1705 & flat\\
        \hline
        \rom{1} & 0.3 & 0.0 & 0.0 & 1.4550 & flat\\
        \rom{1} & 0.3 & 1.25 & 0.0 & 1.4550 & NER\\
        \rom{1} & 0.3 & 1.25 & 0.35 & 1.4530 & flat\\
        \rom{1} & 0.3 & 1.25 & 0.5 & 1.4540 & flat\\
        \rom{1} & 0.3 & 1.75 & 0.0 & 1.4550 & NER\\
        \rom{1} & 0.3 & 1.75 & 0.2 & 1.4526 & NER\\
        \rom{1} & 0.3 & 1.75 & 0.35 & 1.4526 & NER\\
        \rom{1} & 0.3 & 1.75 & 0.5 & 1.4542 & flat\\
        \rom{1} & 0.3 & 2.25 & 0.0 & 1.4550 & NER\\
        \rom{1} & 0.3 & 2.25 & 0.2 & 1.4526 & NER\\
        \rom{1} & 0.3 & 2.25 & 0.6 & 1.4561 & flat\\
        \hline
        \rom{1} & 0.4 & 0.0 & 0.0 & 2.4864 & BH\\
        \rom{1} & 0.4 & 0.025 & 0.0 & 2.4865 & NER\\
        \rom{1} & 0.4 & 0.025 & 0.35 & 2.4795 & NER\\
        \rom{1} & 0.4 & 0.025 & 0.9 & 2.4871 & BH\\
        \hline
        \rom{2} & 0.01 & 0.0 & 0.0 & 1.2431 & BH\\
        \rom{2} & 0.01 & 0.75 & 0.0 & 1.2431 & NER\\
        \rom{2} & 0.01 & 0.75 & 0.35 & 1.2434 & NER\\
        \rom{2} & 0.01 & 0.75 & 0.5 & 1.2445 & NER\\
        \rom{2} & 0.01 & 0.75 & 1.0 & 1.2518 & flat\\
        \hline
        \rom{2} & 0.015 & 0.0 & 0.0 & 2.6532 & BH\\
        \rom{2} & 0.015 & 0.75 & 0.0 & 2.6534 & NER\\
        \rom{2} & 0.015 & 0.75 & 0.35 & 2.6553 & NER\\
        \rom{2} & 0.015 & 0.75 & 1.0 & 2.7000 & BH\\
        \rom{2} & 0.015 & 0.75 & 1.5 & 2.7727 & sBH\\
        \rom{2} & 0.015 & 0.75 & 2 & 2.8812 & sBH\\
        \hline
        \rom{2} & 0.016 & 0.0 & 0.0 & 2.9800 & NER\\
        \rom{2} & 0.016 & 0.75 & 0.0 & 2.9804 & NER\\
        \rom{2} & 0.016 & 0.75 & 0.35 & 2.9829 & NER\\
        \rom{2} & 0.016 & 0.75 & 1.0 & 3.0415 & BH\\
        \rom{2} & 0.016 & 0.75 & 1.5 & 3.1370 & sBH\\
        \rom{2} & 0.016 & 0.75 & 2 & 3.2805 & sBH\\
        \hline
        \rom{2} & 0.02 & 0.0 & 0.0 & 4.3895 & BH\\
    \end{tabular}
    \caption{This table is for $r_0 = 25$, and $w_0 = 6$. $M$ is the initial Misner-Sharp mass, and the outcome indicates the end state of the evolution. It is evident that the addition of the Ricci coupling yields a well-posed IVP.}
    \label{tab:runs}
\end{table}

\bibliography{bibnote.bib}

\end{document}